\documentclass[twocolumn]{aastex631}

\usepackage{xcolor}
\usepackage{graphicx}
\usepackage{float}

\usepackage{hyperref}

\usepackage{multirow}

\begin{document}

\title{The Spin Zone: Synchronously and Asynchronously Rotating Land Planets Have Spectral Differences in Transmission}

\author[0000-0003-3623-7280]{Nicholas Scarsdale}
\affiliation{Department of Astronomy and Astrophysics, University of California, Santa Cruz, CA 95064, USA}

\author{C. E. Harman}
\affiliation{Planetary Systems Branch, Space Science and Astrobiology Division, NASA Ames Research Center, Moffett Field, CA 94035, USA}

\author{Thomas J. Fauchez}
\affiliation{NASA Goddard Space Flight Center, 8800 Greenbelt Road, Greenbelt, MD, 20771, USA}
\affiliation{Integrated Space Science and Technology Institute, Department of Physics, American University, Washington, DC, USA}
\affiliation{NASA GSFC Sellers Exoplanet Environments Collaboration}



\begin{abstract}

New observational facilities are beginning to enable insights into the three-dimensional (3D) nature of exoplanets. Transmission spectroscopy is the most widely used method for characterizing transiting temperate exoplanet's atmospheres, but because it only provides a glimpse of the planet's limb and nightside for a typical orbit, its ability to probe 3D characteristics is still an active area of research. Here, we use the ROCKE-3D general circulation model  to test the impact of synchronization state, a ``low-order'' 3D characteristic previously shown to drive differences in planetary phase curves, on the transmission spectrum of a representative super-Earth land planet across temperate-to-warm instellations (S$_p$=0.8, 1, 1.25, 1.66, 2, 2.5, 3, 4, 4.56~S$_\oplus$). We find that different synchronization states do display differences in their transmission spectra, primarily driven by clouds and humidity, and that the differences shrink or disappear in hotter regimes where water clouds are unable to condense (though our simulations do not consider haze formation). The small size of the feature differences and potential for degeneracy with other properties, like differing water content or atmospheric structure, mean that we do not specifically claim to have identified a single transmission diagnostic for synchronization state, but our results can be used for holistic spectrum interpretation and sample creation, and suggest the need for more modelling in this area.

\end{abstract}



\section{Introduction} \label{sec:intro}


Studies of solar system planets are rich in examples of planets varying in  multiple dimensions on disparate timescales: for example, the Earth's clouds, Mars' storms \citep{SHIRLEY2015128}, Saturn's polar hexagon \citep{Godfrey_1988, SanchezLavega2020}, etc. 
These examples are representative of the way that three-dimensional (3D) effects can shift even observations in one-dimension \citep[e.g., a spatially unresolved spectrum of Earth,][]{Lustig-Yaeger_2023}.
However, although 3D studies of exoplanets have become steadily more common, lower order (typically zero- and one-dimensional) models still predominate in interpreting observations, for two major reasons. 
First, higher dimensional models are computationally expensive to run, especially when investigating a wide parameter space.   
Second, data quality has not always been sufficient to investigate the details of higher dimensional effects. By data quality, we mean not only observational constraining power, but also fundamental physical and chemical materials properties \citep[such as molecular line lists; e.g.,][]{Tennyson_2022} needed to simulate a diverse range of targets. 
Thanks to new instruments and computational facilities, the 3D structure of exoplanets is beginning to be understood in more detail.
For example, recent observations with the James Webb Space Telescope (JWST) and high-resolution ground-based facilities have identified differences between the morning and evening terminators of hot Jupiters \citep[e.g.,][]{Tsai2023, Demangeon2024}. 
As a computational example, atmospheric retrieval analyses have historically relied on 1D models \citep{Benneke_2012}, but new work is pushing towards pseudo-retrievals in 2D and 3D \citep[e.g.,][]{Zingales22_taurex2d, Himes_2023}.
However, there remains much room to explore how three-dimensionality of exoplanets impacts their atmospheres, both from the perspective of underlying physics (and, as an emergent property, habitability), and from our observations of them. 
In particular, 3D studies of planets larger and hotter than Earth, like we simulate here, are less well-represented in the literature than Earth-size, cool-to-temperate planets \citep[e.g.,][]{DelGenio2019_warmearths}.

Outside the domain of hot Jupiters, studies of small exoplanet atmospheres are at the limits of contemporary observatories' capabilities. 
Even the new JWST will require significant investment of observing time to detect primary background gases in terrestrial-type atmospheres \citep[e.g.,][]{Morley_2017_jwst_mstars, Fauchez_2019,Pidhorodetska_2021, Fauchez_2022_thai3}.
This challenge will be all the more significant in light of recent observational evidence that stellar contamination may significantly influence the transmission spectra of small planets around M dwarfs \citep{Moran_2023, Lim_2023}. 
Nonetheless, because of their unique combination of potential habitability and observability, the exoplanet community will spend significant observational time in the search for atmospheres around these planets \citep[e.g., the recent Working Group on Strategic Exoplanet Initiatives recommendation to develop a 500 hour JWST Director's Discretionary Time program for this purpose;][]{2024_ddt_report}.
Forward modelling has an important role to play in informing and interpreting these observations. 

To date, one of the most well-studied characteristics of exoplanets in 3-D is rotation. In particular, synchronous rotators, with permanent day- and night-sides, are thought to be dramatically different from planets with rotation regimes like that of Earth \citep{joshi_2003_synchrotators, merlis_schneider_2010, edson2011, showman2013, Leconte2013, Yang2013, Yang2014, carone2014, Bony2015, noda2017, Haqq-Misra_2018, Zhao2021, lewis2022}.
Planetary synchronization state has a variety of potential effects on planetary atmospheric observability. 
Of particular interest for motivating this work, \citet{Haqq-Misra_2018} showed that there are observably different phase curves as synchronously rotating planets in the habitable zone cross between dynamic regimes. \citet{Wolf2019} also show that several categories of climate regime in the TRAPPIST-1 planets are mutually discernible by phase curve. 
JWST has observed full phase curves of TRAPPIST-1 b and c (Program ID\#3077; PI Gillon) which may provide an observational perspective on this phenomenon when published.
However, phase curves are by definition more time-intensive to obtain than transit or eclipse spectroscopy, and are only feasible for particular systems with sufficient planet-to-star contrast \citep{shporer2017}.
Emission spectroscopy during secondary eclipse is also a useful technique for learning about planetary atmospheres, but becomes relatively more difficult for cooler planets as the planet-to-star flux ratio falls off quickly with temperature \citep[e.g.,][]{Morley_2017_jwst_mstars}.
Therefore, it is of interest to search for the fingerprints of synchronization state in transmission spectroscopy, which can be performed for a wide range of systems.
Along these lines, \citet{Cohen2024} suggest that the rotation rate of synchronously rotating super-Earths can have differing effects on the planet's haze distribution. In particular, they identify three circulation regimes, each with distinct haze distributions and corresponding optical depths. The details of the haze behavior is nonlinear, but they broadly find minima in terminator haze optical depth near rotation periods of $\sim$2 and 15 days, and maxima near $\sim$0.25, 7, and $>$25 days. 
Here, we investigate whether differences caused by rotation can be observed in exoplanet transmission spectra without the effect of significant hazes \citep[which may not dominate the coolest exoplanet atmospheres; e.g.,][]{Yu2021_NatAs, Brande2024, Holmberg2024}. 

We use the general circulation model (GCM) Resolving Orbital and Climate Keys of Earth and Extraterrestrial Environments with Dynamics \citep[ROCKE-3D;][]{Way_2017} to generate a suite of simulations of a representative warm super-Earth exoplanet atmosphere. 
As we describe below, we tailored our simulations to stay outside the conditions notionally associated with a runaway greenhouse scenario, which prevents convergence for our simulations as constructed. 
However, we will argue that some of the fundamental physics seen in our simulations can be generalized to other types of planet. 
In the next section, we describe our modelling setup, along with relevant limitations. In Section \ref{sec:differences}, we present the differences in spectra between our simulations. In Section \ref{sec:discussion}, we investigate the implications our models present for observations of small planets, and provide concluding remarks in Section \ref{sec:conclusions}.

\begin{figure*}[t!]
    \includegraphics[width=0.95\textwidth]{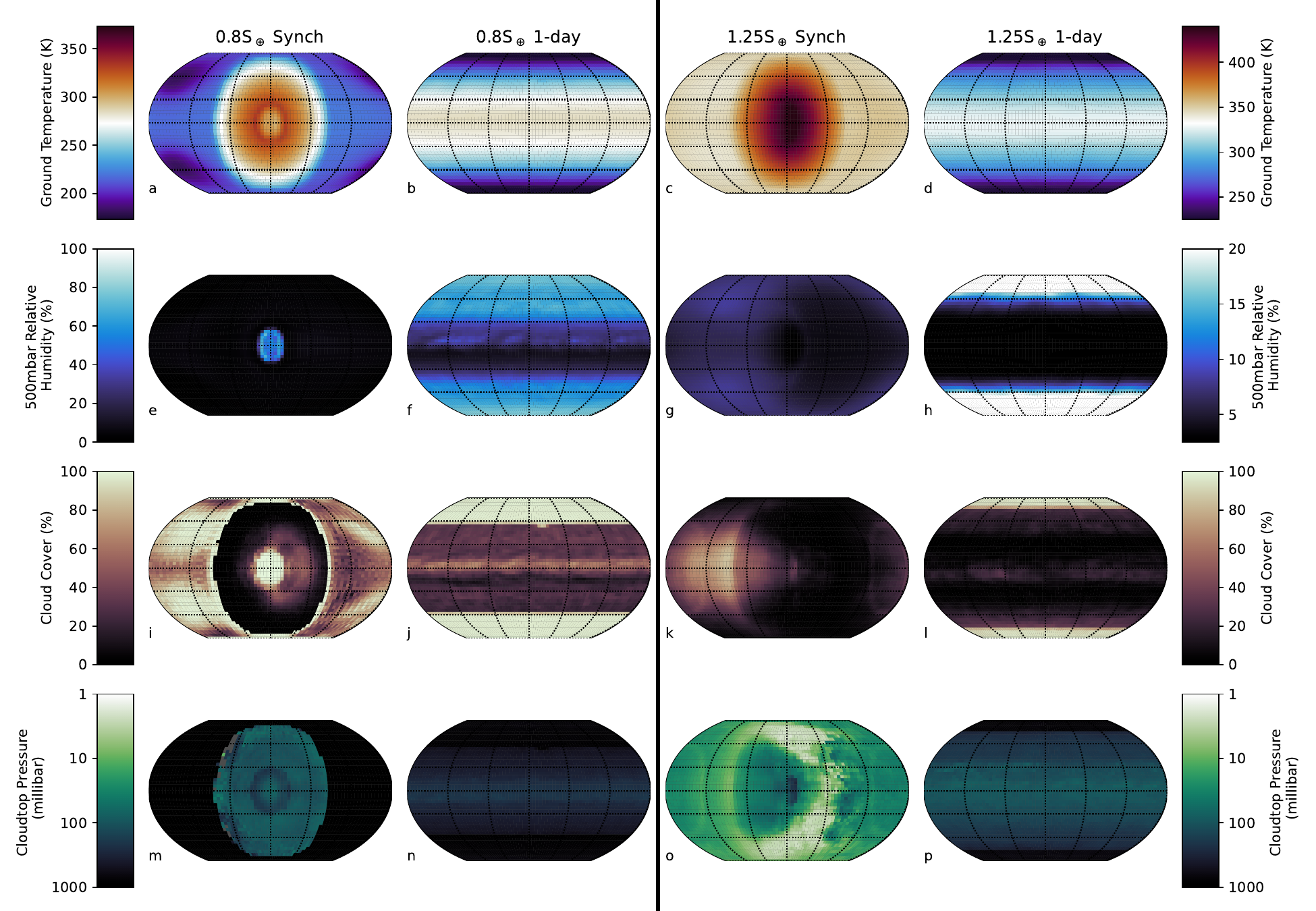}
    \caption{Diagnostic maps of the representative 0.8~S$_\oplus$ and 1.25~S$_\oplus$ simulation instances, with both synchronization states shown. a-d: Ground temperature. Synchronous rotators are dominated by day- to night-side variation, and 1-day rotators by latitudinal variation. e-h: Relative humidity at 500~mbar. i-l: Cloud cover. The cooler simulations are much cloudier. m-p: Cloudtop height in millibars. Note that the left and right halves of the plot use different colorbars.}
\label{fig:tsurf_maps}
\end{figure*}

\begin{figure*}
    \includegraphics[width=0.95\textwidth]{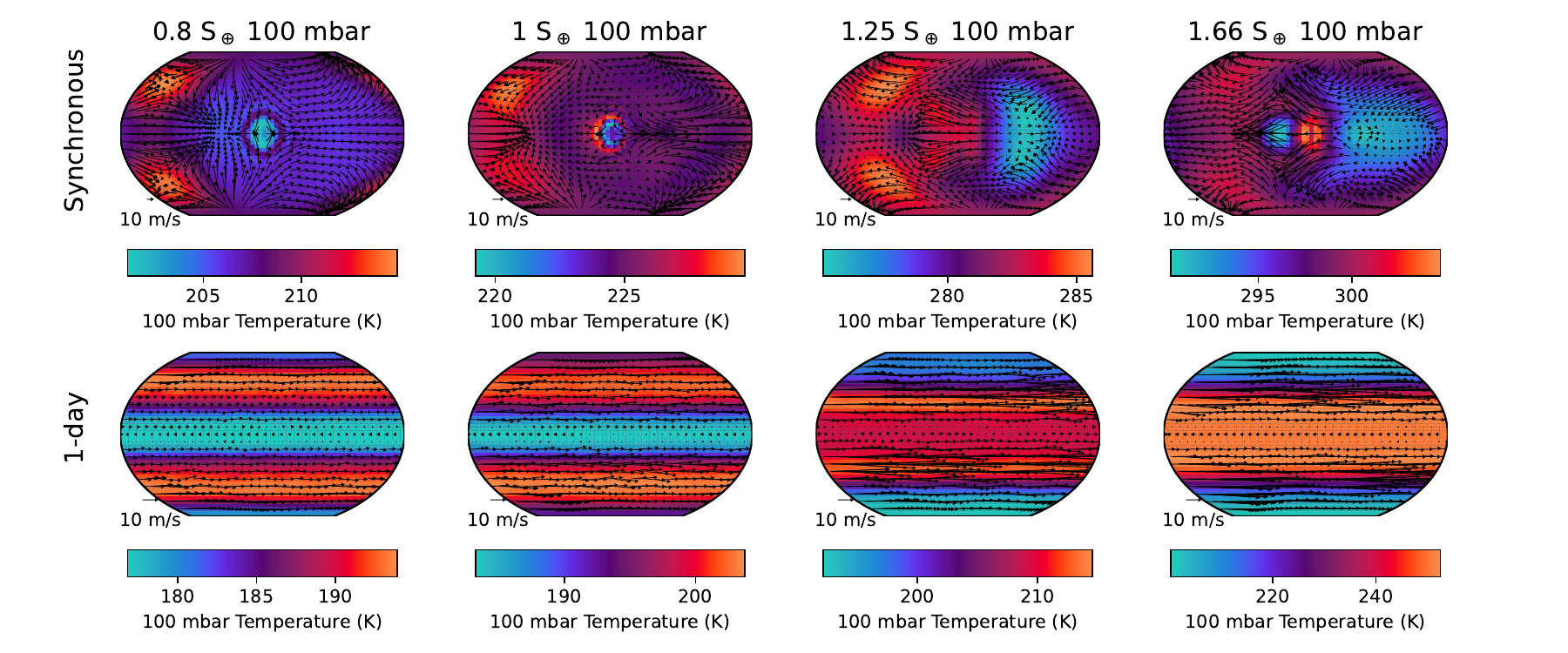}
    \caption{Slices through the four coolest simulation atmospheres at 100mbar. Top: Synchronous rotators. Bottom: 1-day rotators. }
\label{fig:winds_temp_100mbar}
\end{figure*}

\section{Simulation Details} \label{sec:methods}

\subsection{Experimental Setup}

ROCKE-3D is a planetary general circulation model developed from (and in parallel with) the NASA Goddard Institute of Space Studies (GISS) ModelE2, and Earth climate model \citep{Schmidt2014}. ROCKE-3D has been used to simulate terrestrial, exoplanet, and Martian climate \citep[e.g.,][]{DelGenio2019, Guzewich2021}, including for planets at higher instellations and in different synchronization states \citep[e.g.,][]{Fujii2017, Kane2018, Colose2021, He2022}. More details about ROCKE-3D can be found in \citet{Way_2017}. We make use of the Planet 1.0 version of the model, which includes the SOCRATES radiation scheme \citep{Edwards1996, Edwards1996b}. 

We created a series of three-dimensional simulations of a terrestrial exoplanet atmosphere. Our model uses the mass and radius of L 98-59 d \citep{Kostov_2019_l9859, Cloutier_2019_l9859}, a super-Earth sized planet that has been identified as a promising target for transmission spectroscopy \citep{Pidhorodetska_2021}. We choose the stellar spectrum for Kepler-1649, an M dwarf with an effective temperature of 3240~K \citep{Angelo2017}, which was the closest match available for the warmer L 98-59 \citep[T$_{eff}=$3412~K;][]{Cloutier_2019_l9859}. We scale the total flux to match L 98-59 so that the planet receives the targeted irradiation from our simulation grid. Our simulations vary synchronization state (P$_\textrm{rot}=1$~d, P$_\textrm{orb}$) and irradiation (S$_p=$0.8, 1, 1.25, 1.66, 2, 2.5, 3, 4, 4.56~S$_\oplus$), in order to search for differences that might be detectable by transmission spectroscopy. Additional details about our simulation can be found in Table \ref{tab:params}. We additionally run three 3:2 P$_{rot}$:P$_\textrm{orb}$ cases (at 1, 1.25, and 1.66~S$_\oplus$ and find that they qualitatively resemble the 1-day rotators with less extreme pole-to-equator temperature differences. We are motivated to vary synchronization state because, although the circularization timescale is often taken to be short (e.g., \citealt{KASTING1993108}, following \citealt{Peale1977}), recent work has shown that the dynamical state of small planets is complex (particularly in multi-planet systems like L 98-59, e.g., \citealt{vinson_2019_spinstates}) and may induce spin very unlike the classical picture of a synchronous rotator, with a permanent day- and night-side. 
These differing dynamical states will lead to climatological differences \citep[e.g.,][]{Shakespeare2023, Chen2023}. Because a complete dynamical simulation tied to a GCM is computationally unsuited for a parameter space sweep, we use our 1-day rotators to represent both planets which have not yet fully circularized, and planets which have been excited from their synchronous orbits by mutual dynamical interaction, in that they broadly have more efficient heat redistribution. We emphasize, however, that the climate effects of the latter state may be highly complex and time-variant, and our models here serve as end-member cases for latitudinal (1-day) and sub-/anti-stellar (synchronous) energy transport.
Our simulations have zero obliquity and eccentricity, both parameters shown to influence climatology in comparable ways to rotation \citep[e.g.,][]{Colose2021, Olson_2020, He2022, Jernigan_2023}, although given the potential for dynamically complex states, these would be worthwhile to investigate in future work. 

The simulations presented here are run at 4$^\circ\times5^\circ$ latitude-longitude resolution with 40 atmospheric layers from the surface to $\sim$0.1 mbar at the top of the model domain. 
The surface is initialized as a fully saturated sand, which represents the entire liquid water inventory as we have opted not to run with a large surface water inventory. This helps avoid a transition to a runaway greenhouse scenario \citep{Goldblatt2013}. 
We make this choice not because a significant water inventory is physically implausible, but because it cannot be investigated with our model. The runaway greenhouse scenario that results on hot terrestrials with large water inventories has only recently been examined with a GCM model for the first time \citep{chaverot2023_runaway, Turbet2023}.
The presence of oceans introduces additional climate degeneracy driven by the distribution of land and water \citep{Macdonald2022, Macdonald2024}, motivating our choice to focus on land planets.
However, we also note that the assumption of low water content does have some physical motivation, since 
planets with M-type host stars (like L 98-59) may lose significant amounts of their starting water content due to the host's long pre-main-sequence phase \citep[e.g.,][]{lugerbarnes_2015_waterloss}. 

\begin{table}
    \centering
    \begin{tabular}{|l|c|c|c|}
    \hline
        Simulation & P$_\textrm{orb}$ (d) & P$_\textrm{rot}$ (d) & Instellation (S$_\oplus$)\\
    \hline
    \hline
        0.8\_1day & 27.285 & 1 & 0.8\\
        0.8\_synch & 27.285 & 27.285 & 0.8\\
    \hline
        1\_1day & 23.250 & 1 & 1\\
        1\_synch & 23.250 & 23.250 & 0.8\\
    \hline
        1.25\_1day & 19.667 & 1  & 1.25\\
        1.25\_synch & 19.667 & 19.667 & 1.25\\
    \hline
        1.66\_1day & 15.898 & 1  & 1.66\\
        1.66\_synch & 15.898 & 15.898  & 1.66\\
    \hline
        2\_1day & 13.824 & 1 & 2\\
        2\_synch & 13.824 & 13.824  & 2\\
    \hline
        2.5\_1day & 11.694 & 1  & 2.5\\
        2.5\_synch & 11.694 & 11.694 & 2.5\\
    \hline
        3\_1day & 10.200 & 1  & 3\\
        3\_synch & 10.200 & 10.200  & 3\\
    \hline
        4\_1day & 8.220 & 1  & 4\\
        4\_synch & 8.220 & 8.220 & 4\\
    \hline
        4.56\_1day & 7.4507 & 1  & 4.56\\
        4.56\_synch & 7.4507 & 7.4507  & 4.56\\
    \hline
    \end{tabular}
    \caption{The orbital and rotational periods, irradiation values, and rotation states for each of our simulations.}
    \label{tab:periods}
\end{table}

\begin{table}
    \centering
    \begin{tabular}{|l|c|c|}
    \hline
        Property & Units & Value \\
    \hline
        Stellar Spectrum & & Kepler-1649 (3240 K) \\
        Stellar T$_{eff}$ & K & 3412 \\
        Planet Radius & km & 9690 \\
        Planet Surface Gravity & m~s$^{-2}$ & 8.23 \\
        Obliquity & & 0 \\
        Eccentricity & & 0 \\
        \multirow{2}*{Composition} & & 99\% N$_2$, 400ppm CO$_2$,\\
         & &  variable H$_2$O\\
    \hline
    \end{tabular}
    \caption{The planet and atmospheric parameters used for our simulations.}
    \label{tab:params}
\end{table}

The simulations were run in batches, starting with the lowest instellation cases. The outputs of these simulations were then used as initial conditions for the higher instellation cases, as they were closer to the expected final surface temperatures than the default Earth-like input files. We found that this was necessary to achieve convergence in the hotter simulations. To rule out hysteresis, we reran the 1~S$_\oplus$ synchronous simulation using the 3~S$_\oplus$ synchronous case as its initial condition, and found that the resulting climate state was qualitatively indistinguishable. Additionally, our models lack oceans and except in the coolest two cases have no surface ice, eliminating those drivers of contrast in the planetary albedo. Previous studies also find that climate hysteresis in terrestrial atmospheres is reduced for planets with M-type host stars \citep{Shields2014} and tidally locked planets \citep{Checlair2017, Checlair2019}. 

We provide diagnostic maps of surface temperature for two representative irradiation cases in Figure \ref{fig:tsurf_maps} (top row), wind velocity at 100~mbar as a dynamics diagnostic in Figure \ref{fig:winds_temp_100mbar}, and substellar pressure-temperature profiles in Figure \ref{fig:convergediags}a. Convergence to a steady state climate is estimated from the time derivative of surface temperature and imbalance between the incoming and outgoing radiation fluxes (Figure \ref{fig:convergediags}b and c, respectively), both of which should be near zero. In our cases, we found that $<1$ W/m$^{2}$ variations and corresponding temperature fluctuations were common. We attribute this to the limited water inventory and moderate timescale for equilibration with the atmosphere 
in all but the hottest cases (see Section \ref{subsec:limitations}).

Once the simulations were converged, we used the Planetary Spectrum Generator \citep[PSG,][]{villanueva2018_psg,villanueva2022_psg_handbook}  in order to obtain transmission spectra. PSG and the associated \texttt{GlobES} module\footnote{https://github.com/nasapsg/globes} uses the full three-dimensional output of the GCM to self-consistently perform the 3D radiative transfer calculations of the spectrum in transmission. Because ROCKE-3D only generates atmospheres up to pressures of 0.1~mbar and lower pressures can still impact transmission spectroscopy, we extended the atmospheres up to $\sim$10~nbar, assuming an isothermal profile and the same molecular mixing ratios, mimicking the procedure in \citet{Fauchez_2022_thai3} (and similarly finding minimal differences). We also applied an additional pre-processor option to obtain the cloud mixing ratio and particle size variables for the liquid and water ice clouds. 
We followed the recommendations in the PSG handbook \citep{villanueva2022_psg_handbook} for generating transmission spectra from the .aij* ROCKE3-D output files to ensure that the correct geometry was captured in our spectra.

\subsection{The Limits of ROCKE-3D}
\label{subsec:limitations}

ROCKE-3D was originally developed as a branch of, and in parallel with, Earth climate simulation capabilities \citep[ModelE2,][]{Schmidt2014}. As a result, it is best suited for modelling small, temperate planets. The parameters of the planet we chose to model are at the limits (and in some cases beyond the limits) of where ROCKE-3D and its underlying correlated-k tables are appropriate given our model setup. 
We therefore provide here a list of limitations of our model. 
Only recently has a different publicly available GCM tool that can bypass some of these limitations (particularly those related to the runaway greenhouse effect) been developed \citep{chaverot2023_runaway, Turbet2023}.

\begin{figure*}[t!]

\includegraphics[width=0.99\textwidth]{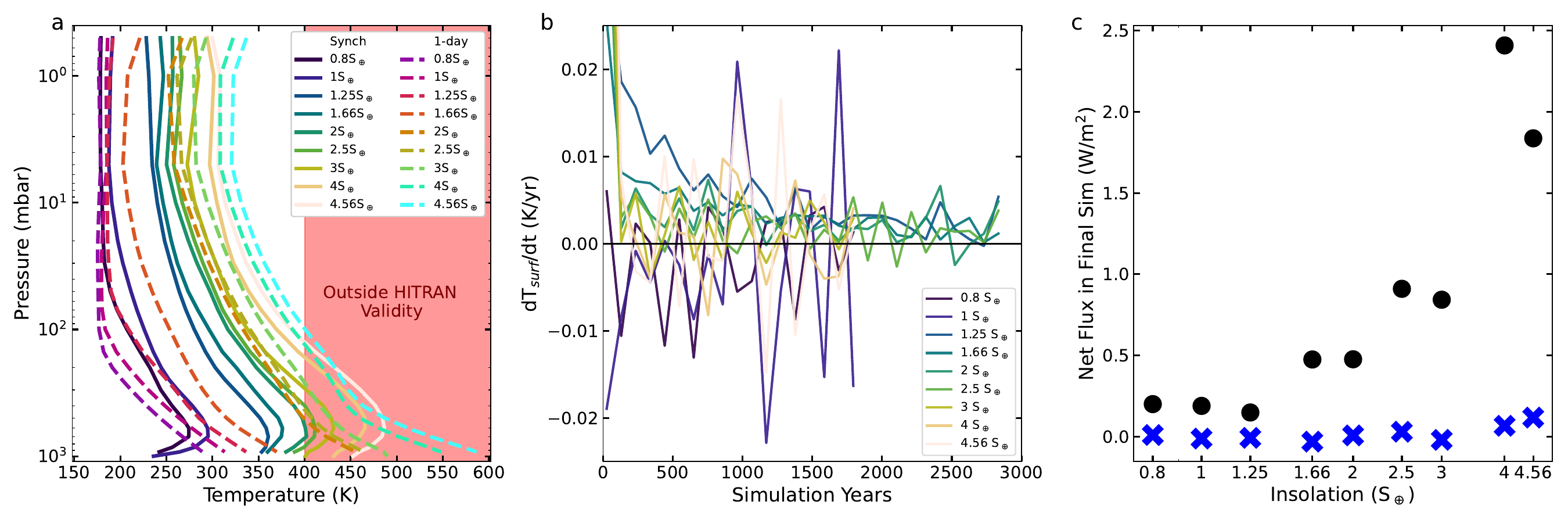}
\caption{Convergence diagnostics for our simulations. a: The pressure-temperature profiles of each simulation at the sub-stellar point (or longitude=0 for the 1-day rotators). The region where the HITRAN line list is inadequate is highlighted in red. b: The average surface temperature at each time point of our synchronously rotating models, shown for each irradiation step.
c: The energy balance of each simulation. A value of 0 would indicate a perfectly equilibrated simulation. 
The hottest synchronous rotators display non-zero net fluxes. But, values lower than $\sim$0.75~W~m$^{-2}$ are sufficient to be considered converged (although see Section \ref{subsec:limitations} next).
}
\label{fig:convergediags}
\end{figure*}

\begin{enumerate}
    \item Small exoplanets are thought to have a variety of bulk compositions, leading to great diversity in their atmospheres \citep[see, e.g.,][]{Wordsworth2022}. However, to maintain relative simplicity, our realization of this planet is narrow. Our model is a dry, ``desert'' world (i.e., no oceans). It is decidedly not a water world \citep[which this planet may be, per][]{luque_22_waterworlds}.
    Its atmospheric composition is entirely N$_2$/CO$_2$ with trace H$_2$O. It also does not include significant topography. 
    \item Our compositional choices result in an atmosphere that does not have significant haze particles, photochemical or otherwise. Such hazes may be common in warm exoplanet atmospheres \citep[e.g.,][]{Gao_2021_aerosols_review}, but are excluded here to limit the project's scope. Depending on haze properties, they have been shown to significantly alter planetary climate for larger exoplanets \citep[e.g.,][]{Steinrueck2023} and therefore significantly expand the parameter space to be explored.
    We point the reader to \citet{Cohen2024} for a GCM-driven examination of hazes' impact on small planet transmission spectra, and \citet{Fauchez_2019} for a comparison of the effects of clouds and hazes. 
    \item Some of our hotter simulations, in parts of their atmosphere, exceed the limits at which the HITRAN molecular line list for H$_2$O is valid (see Figure \ref{fig:convergediags}a), which results in problems for the radiative transfer code \citep{Goldblatt2013}. We include them here for completeness. Given our model setup, the physics of these hotter simulations are qualitatively sensible, since in both rotation regimes, they converge on a perpetually sub-saturated atmosphere as the temperature increases. These models produce spectra that resemble the higher-temperature simulations that are within the valid HITRAN temperature range. However, we cannot be confident in results outside the valid HITRAN range; thus, the bulk of our conclusions are drawn from the cooler planets, where no line-list issues exist. 
    \item We self-consistently calculate the orbital distance and rotation rate for each instance, meaning that the synchronous rotators approach the dynamical regime sampled by the 1-day rotators in rotation period as we move to higher instellation (at a 1 day orbital period they would have the same rotation period). Our hottest simulation has an orbital period of $\sim$7 days, which likely crosses into the Rhines rotation regime \citep{Haqq-Misra_2018}. However, per the previous point, there is limited information to be gained from the hotter simulations, and the cooler cases resemble the slow rotators (Figure \ref{fig:tsurf_maps}).
    \item Our simulations are largely within the net flux threshold for convergence ($\lesssim$1~W~m$^{-2}$ deviation from 0), and for the most part their temperatures do not fluctuate significantly in the final stages of the model, which is a useful stability indicator (Figure \ref{fig:convergediags}b). However, particularly in the hottest synchronous cases, there is non-zero net flux (see Figure \ref{fig:convergediags}c) and persistent temperature fluctuation at the 1-2~K level. Even though the hottest cases still broadly appear converged, there are clear limitations on what information can be extracted from them, and we include them here primarily for completeness. 
\end{enumerate}


\section{Some Synchronous and 1-Day Rotators Have Different Spectra} \label{sec:differences}

\begin{figure*}[t!]
\includegraphics[width=0.95\textwidth]{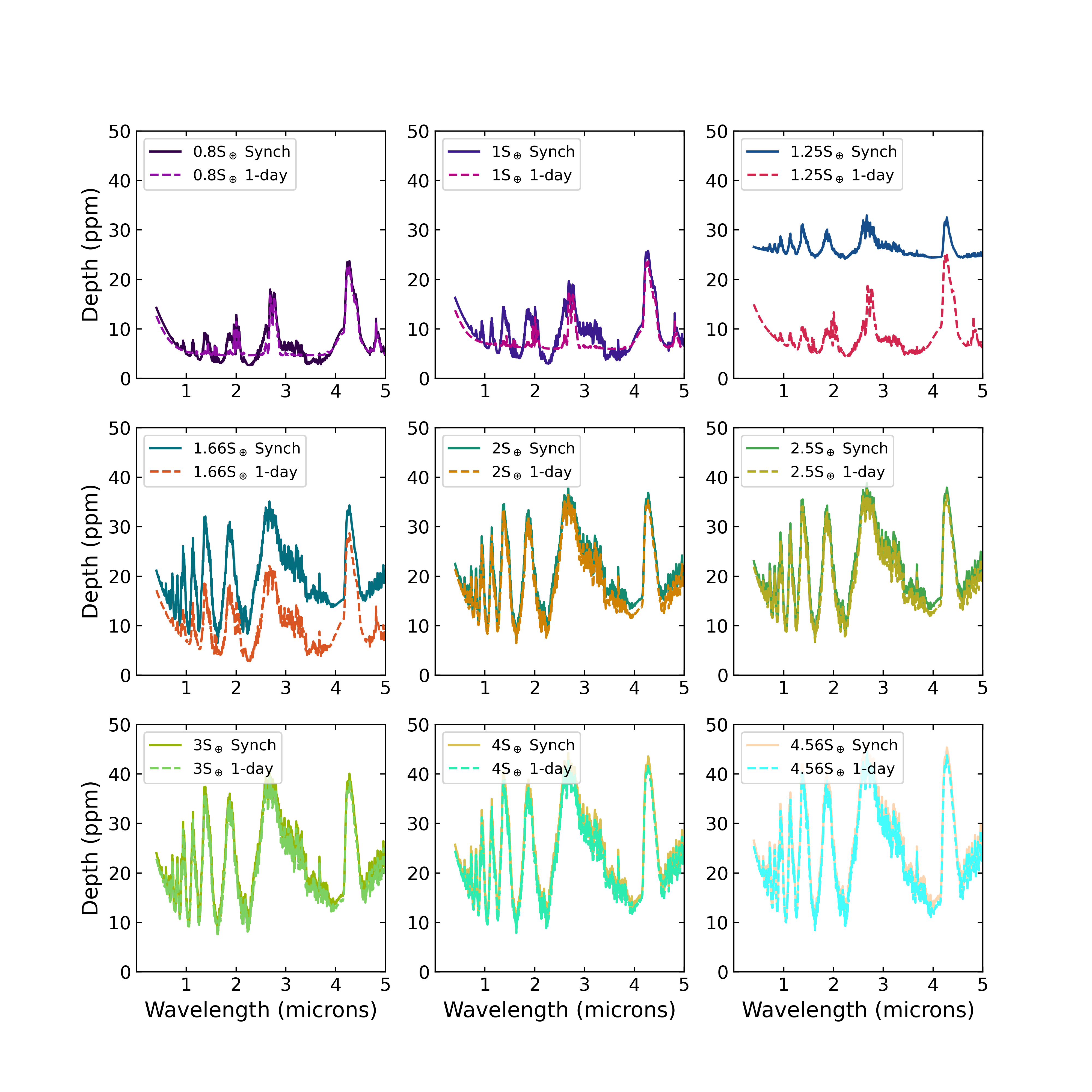}
\caption{The spectrum of each simulation in our explored parameter space. Each has the same composition (N$_2$-dominated, trace CO$_2$ and H$_2$O; Table \ref{tab:params}). Each box shows the synchronous and 1-day rotator cases for different instellation values, which are printed in the legend. Up to 1.66~S$_\oplus$, the transit spectra are distinguishable. In warmer cases, the differences disappear.}
\label{fig:spectra}
\end{figure*}

The primary result of our work can be summarized as follows: for planets from roughly temperate-to-Venus levels of irradiation (1-2~S$_\oplus$), the near-infrared transmission spectra of planets in the two rotation regimes we consider are visibly distinct. This is clear in Figure \ref{fig:spectra}, where we show the synchronous vs 1-day rotator transmission spectra for each model in our instellation grid. Note that this representation differs slightly from the usual representation of transmission spectra. Rather than showing total depth (i.e., including the entire planetary radius), or showing only feature height above the opaque planet + continuum (i.e., minimum of feature height in all plots is 0), we show feature amplitude above the planet surface only. Thus the opaque atmosphere also contributes to the continuum, resulting in ``depths'' that have minima $\gtrsim$3ppm. 
We have selected the 0.5-5 micron wavelength region specifically because this is where JWST will be able to obtain precise spectra for small planets. 

In the simulations up to roughly Venus irradiation, the spectra are distinct in terms of their spectral continuum and features, amplitude of features, and scattering slope, given a spectrum of arbitrarily high precision (see Section \ref{sec:observations} for a discussion of the implications of these differences for real observations). In the following subsections, we will discuss the characteristics of these differences and the factors that drive them in more detail. We discuss clouds first and then climate factors. We note that these factors are of course intrinsically linked, but are frequently measured separately in atmospheric retrievals, where a deck of grey clouds at arbitrary altitude is a common tool used to match observations \citep[e.g.,][]{Welbanks_2021}. 

\subsection{Differences I: Clouds}\label{sec:diffclouds}

ROCKE-3D, and therefore our simulations, considers water (liquid and ice) clouds only. The full treatment of clouds in ROCKE-3D is described in \citet{Way_2017}, Section 4.2. To briefly summarize, the simulation treats the dynamics, thermodynamics, and microphysics of convection for a mass flux of rising moist air \citep{DelGenio2015}. The simulation maintains sub-saturated grid conditions (in contrast to some other GCMs) in cells where updrafts originate. The sub-gridbox treatment allows nonzero cloud fractions to appear without full saturation of a grid cell and are instead determined from local humidity and stability \citep[see][]{DelGenio1996, Schmidt2006, Schmidt2014}. We refer the reader to \citet{Way_2017} for the full model details.

Clouds are a well-known and significant driver of the shapes of exoplanet spectra \citep[e.g.,][]{Kreidberg2014, Knutson2014, Fauchez_2019, Komacek2020, Suissa2020, Gao_2021_aerosols_review, Grant2023_tstdreams}. Our simulations are no different: many spectral differences that we show here are clearly driven by clouds. This is evident by inspection, particularly in the coolest cases where a flat (i.e., spectrally grey) continuum opacity characteristic of clouds is present (top row, Figure \ref{fig:spectra}). 

\begin{figure*}[t]
\includegraphics[width=0.95\textwidth]{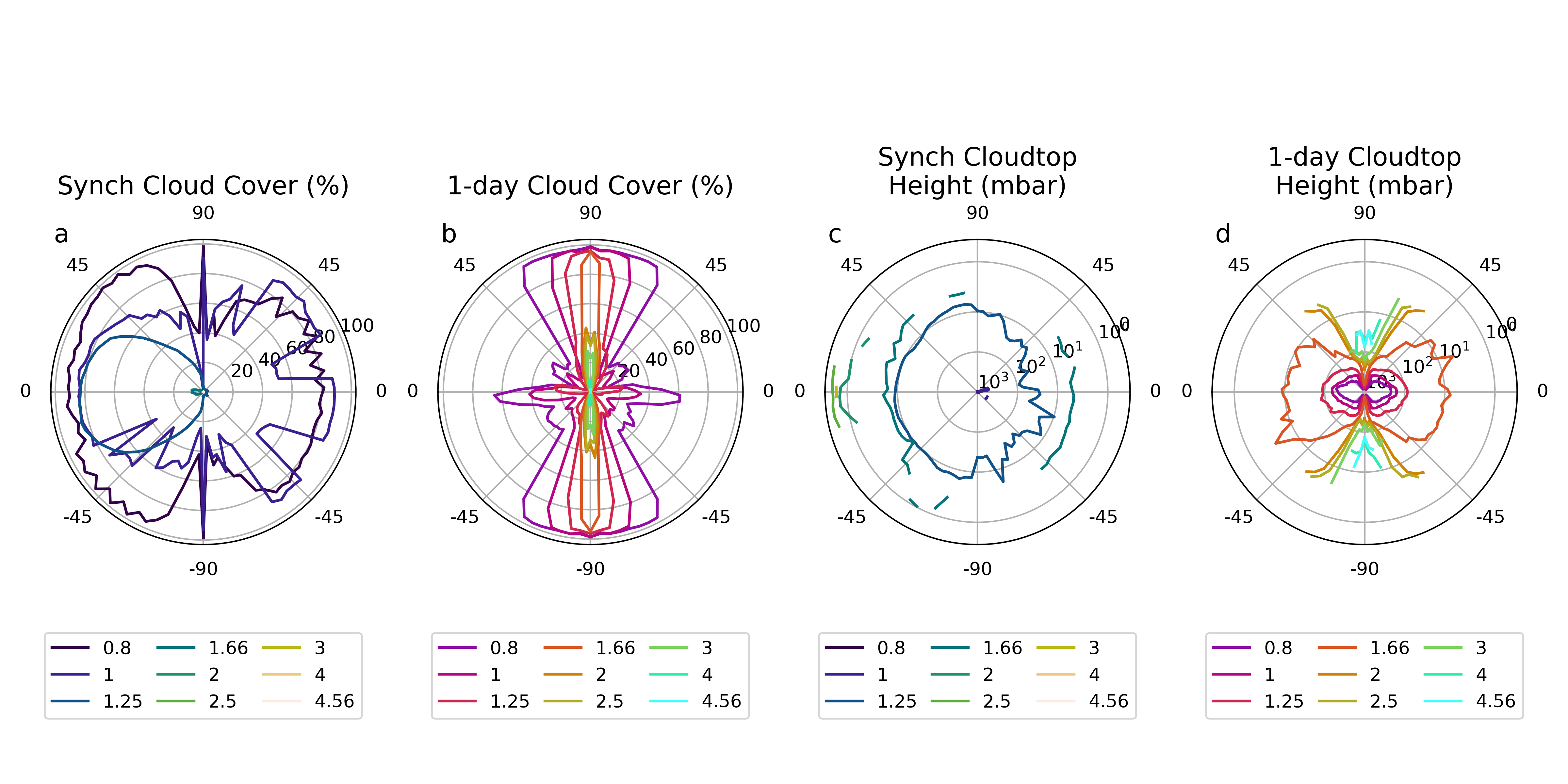}
\caption{The terminator cloud properties of our simulations as a function of latitude. The left and right sides correspond to the morning and evening terminators, respectively. a-b: The fractional cloud cover of each cell for the synchronous (a) and 1-day (b) rotators. c-d: The height of the cloudtop for the synchronous (c) and 1-day (d) rotators, in millibars. Higher-pressure (lower-altitude) clouds are radially further from the center.}
\label{fig:clouds}
\end{figure*}

These continuum differences become less pronounced and eventually disappear in both rotation cases once we raise the instellation as high as 2~S$_\oplus$. This occurs because the hotter simulations have little to no cloud cover, and in a cloudy atmosphere, the continuum level is defined by the altitude of the optically thick cloud deck. In Figure \ref{fig:clouds}a and b, we show the total cloud cover in each simulation's terminator region, which drives the transmission spectrum. For the synchronous rotators, once instellation rises to 1.66~S$_\oplus$ case, cloud cover becomes very low, and in the hotter cases still, there is negligible cloud cover. This means that even though the terminator clouds in the hotter simulations rise to higher altitudes and would correspondingly ``mute'' more of the transmission spectrum by obscuring a larger fraction of the atmosphere in the vertical direction (Figure \ref{fig:clouds}b and c), their coverage is so low that the muting is imperceptible. In all cases, the cloud cover is almost exclusively stratiform. Convective cloud cover is negligible except in the coolest case (0.8~S$_\oplus$), where it makes up a few \% of the total cloud coverage.

Broadly, this result is in agreement with \citet{Kopparapu2016}, who find as they increase the instellation of their synchronous rotators, this drives the water vapor greenhouse, increasing absorption at high altitudes. This in turn decreases the lapse rate and suppresses convection, muting cloudiness. The 1-day rotators still retain appreciable cloud cover in the 1.66~S$_\oplus$ case, being cooler than the synchronous rotators at the same irradiation (Figure \ref{fig:convergediags}a), but follow the same trend of moving to lower and eventually no cloud cover as instellation increases. 

\begin{figure}[h]
\includegraphics[width=0.45\textwidth]{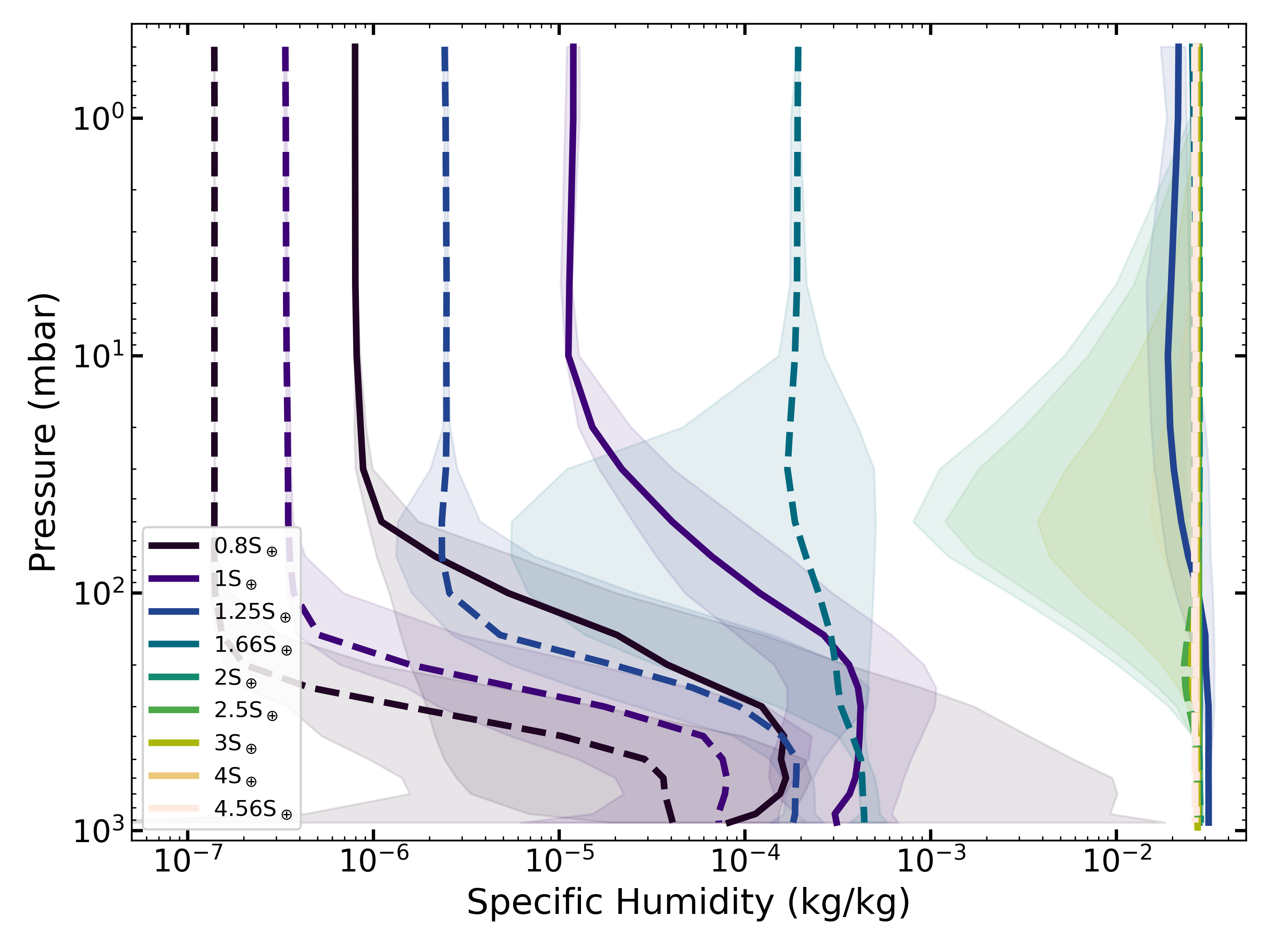}
\caption{The mean pressure-humidity profile for each simulation. Envelopes span the minimum and maximum specific humidity at each pressure level. Unlike in other figures, the color scheme is the same for the synchronous and 1-day cases to enable easier comparison between instellations, and the two synchronization states are indicated by line style (solid lines for synchronous, dashed lines for 1-day).The transition to nearly sub-saturated atmospheres occurs at lower irradiation for the synchronous rotators than the 1-day rotators.  
}
\label{fig:humidity_125}
\end{figure}

To understand why this is the case, we show the specific humidity profiles in Figure \ref{fig:humidity_125}: by the time the instellation reaches 2~S$_\oplus$, both rotation regimes have converged on a perpetually sub-saturated atmosphere. The transition occurs at different irradiation thresholds for the synchronous and 1-day cases because of their different climates - the synchronously rotating cases are hotter than the 1-day rotators in the intermediate instellation cases (Figure \ref{fig:convergediags}a). Note in the 1.66~S$_\oplus$ case (Figure \ref{fig:spectra}, middle left panel), the  synchronous rotator's spectrum has already begun to look identical to the hottest cases, but not so for the 1-day rotator. 

Why do the highest-instellation cases have roughly identical spectra (modulo a small offset), regardless of synchronization state? First, we caution the reader that in these cases (certainly by 2.5~S$_\oplus$), we have begun to enter a regime of questionable validity of the HITRAN line lists \citep[Figure \ref{fig:convergediags} a;][]{Goldblatt2013}, so these results should be treated with caution. However, the behavior can still be interpreted qualitatively. The only spectrally active characteristics of our model atmospheres are CO$_2$, H$_2$O, N$_2$-N$_2$ collision-induced absorption (CIA), and clouds. Without clouds and their potential associated climate feedback, the hottest cases are composed of perpetually sub-saturated water, N$_2$, and trace CO$_2$, whose medium resolution spectral signatures are not changing significantly with temperature. 
The amplitude of spectral features does increase slightly with instellation, as expected given the temperature dependence of the atmospheric scale height. 

Remaining for discussion is the 1.25~S$_\oplus$ case, in which the synchronous rotator displays apparently anomalous behavior in its transmission spectrum, relative to the other simulations. The transit depth above the defined surface (i.e., the continuum height) is consistently $\sim$25 ppm, whereas the other simulations range from 2-10 ppm from coolest to warmest. The reason for this appears to be significantly higher-altitude clouds than in any of the cooler simulations, which we show in Figure \ref{fig:clouds}c. The 1.25~S$_\oplus$ case synchronous rotator has terminator clouds more than an order of magnitude higher in the atmosphere than either its cooler synchronously rotating counterparts or its 1-day rotator analog, resulting in this ``lifted'' continuum. Although the 1.66~S$_\oplus$ synchronous case has even higher altitude clouds, their coverage is extremely low (Figure \ref{fig:clouds}a), so they do not contribute significantly to the opacity. This can also be seen comparing the corresponding panels in Figures \ref{fig:spectra} and \ref{fig:spectra_cloudfree}, where eliminating the clouds minimally changes the 1-day rotator's spectrum but dramatically affects the synchronous rotator's. 
From a dynamics perspective, this simulation's high altitude clouds appear to be driven by dayside winds carrying warmer, drier air crashing into the cooler, wetter equatorial morning terminator (Figure \ref{fig:winds_temp_100mbar}).

\subsection{Differences II: P-T Profiles and Humidity}\label{sec:diffptprofiles}

\begin{figure*}[t!]
\includegraphics[width=0.95\textwidth]{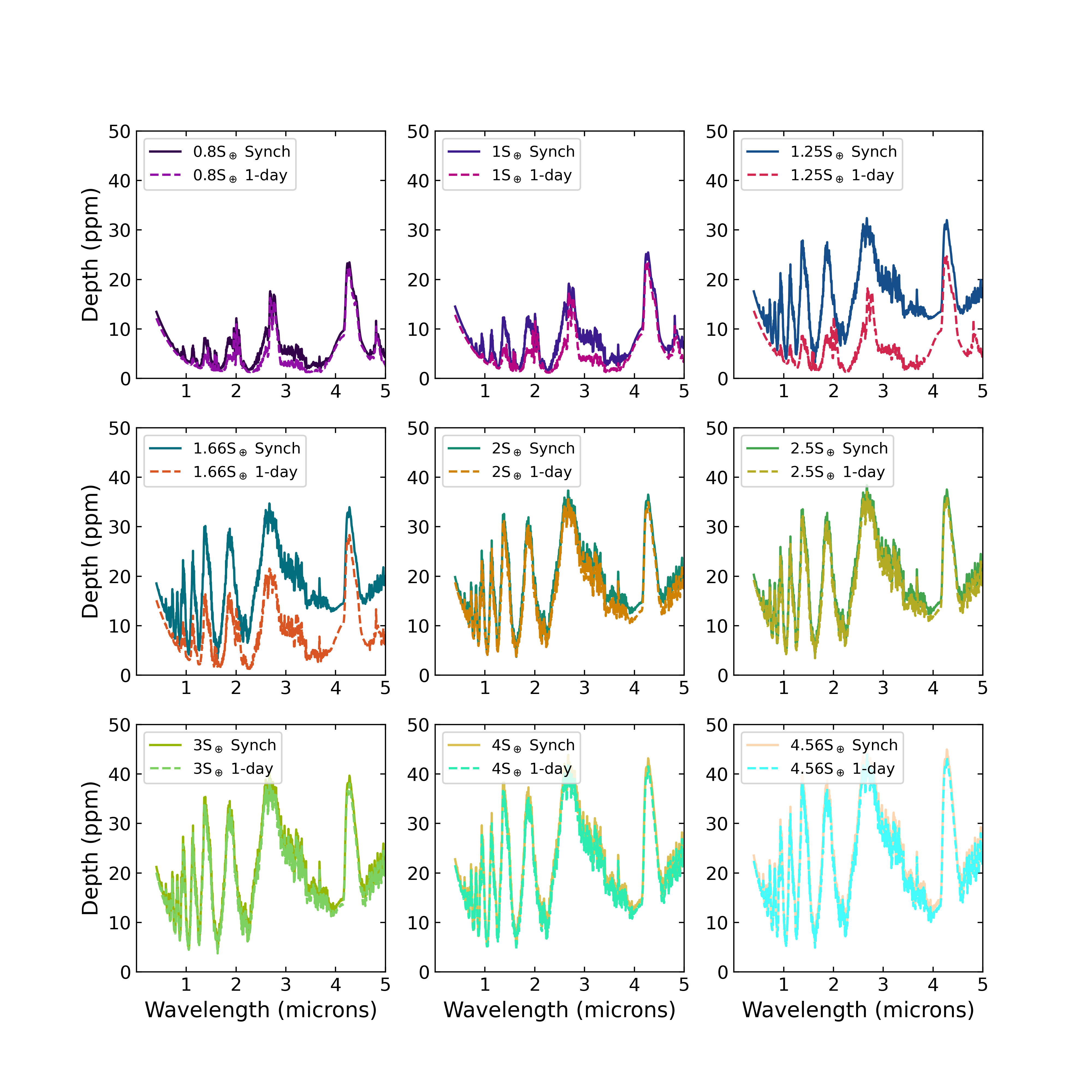}
\caption{The spectra of each instantiation of our simulation with the optical and spectral effects of clouds removed. As in Figure \ref{fig:spectra}, each box shows the synchronous and 1-day rotator cases for different instellation values, which are printed in the legend. Despite the absence of clouds, visible (though smaller) differences persist in the spectra of less-irradiated models.} 
\label{fig:spectra_cloudfree}
\end{figure*}

Clouds are a significant driver of the differences in our spectra, but not the only one. In order to test this, we repeated our procedure for generating the spectra with \texttt{globes} and PSG, but with clouds disabled. The results can be seen in Figure \ref{fig:spectra_cloudfree}, where differences persist in some of our spectra despite the absence of clouds when calculating the transmission spectrum. We emphasize that in this case, we are only ``turning off'' clouds in the spectrum calculation, but the resulting spectrum is created from a simulation whose climatology was sculpted by clouds. This exercise is only a diagnostic of what transmission spectrum feature differences are directly caused by clouds. Qualitatively, the continuum absorption is more similar between rotators in the cloud-free cases (as expected), but some feature heights remain different. In particular, the 1.25 and 1.66~S$_\oplus$ cases show significant differences in the amplitude of the water features from 1-3~$\mu$m, even in the spectral absence of clouds. As before, in the warmest cases the spectra are indistinguishable.

What is driving these persistent differences? Given the relative simplicity of our models, the atmospheric pressure-temperature and humidity profiles are the two prime suspects. The atmospheric $\tau=1$ surface will change as the water abundance varies in the atmosphere, represented by these two factors.  
The 1.25~S$_\oplus$ case, along with the 1.66~S$_\oplus$ case, have the most qualitatively significant differences between the two rotators' cloud-free spectra. Revisiting the pressure-temperature profiles in Figure \ref{fig:convergediags}a, we can see that the temperature difference between synchronous and 1-day rotators is largest in these two cases. This will of course drive some differences in the scale height. However, more significantly, these temperature differences are driving the synchronous rotators in these two cases to reach a just sub-saturated state, while the atmospheres of the 1-day rotators are drier (Figure \ref{fig:humidity_125}). This is likely the dominant driver of spectral differences in the cloud-free cases.  



Driven by the pressure-temperature profile, the water content of the atmosphere also differs significantly between these cases (Figure \ref{fig:humidity_125}). In the 1.25~S$_\oplus$ case, the hotter synchronous rotator has already moved much of its surface water into atmospheric humidity, where the 1-day rotator remains fairly dry across all latitudes. Tracing the specific humidity across instellation for each rotation case, there is a clear transition from relatively dry to sub-saturated atmospheres. However, the break point occurs earlier (i.e, at lower instellation) for the synchronous rotators than the 1-day. The higher global temperatures (at least in intermediate cases) of the synchronous rotators are the likely driver of this. 

\section{Discussion}
\label{sec:discussion}

\subsection{Comparison To Previous Work}
\label{sec:comparison}

Before discussing the observational characteristics of our simulations, we continue the thread of Sections \ref{sec:diffclouds} and \ref{sec:diffptprofiles} and compare the climatology of our simulations to previous works which investigated similar scenarios. Much work on super-Earths has focused on water worlds \citep[e.g.,][]{Yang2013, Sergeev2022_thai2}. We selected the Hab 1 case \citep{Sergeev2022_thai2} from the TRAPPIST-1 Habitable Atmosphere Intercomparison (THAI) series of papers \citep{Turbet2022_thai1, Sergeev2022_thai2, Fauchez_2022_thai3} as a ``standard'' water world suitable for comparison to our simulations. Both have an N$_2$-dominated atmosphere with 400ppm CO$_2$, although our simulations are much drier: the ROCKE-3D implementation of Hab 1 has a 90-meter global ocean, compared to our simulations whose only water is in the moist soil. Hab 1 is also slightly less irradiated than our coolest simulation (900~W~m$^{-2}$ vs 1089~W~m$^{-2}$ in our 0.8 ~S$_\oplus$ case). 

Our simulations' behavior differs from that of the THAI Hab 1 case, which has substantial cloud cover at the substellar point frequently seen in synchronous rotators in 3-D \citep[e.g.,][]{Yang2013,Yang2014,Kopparapu2016}. Instead, our simulations qualitatively better resemble the water-limited cases explored in \citet{Lobo_2023}, with less dayside cloud cover and correspondingly lower dayside albedo. We find similarly large ($\gtrsim$100~K) day-to-night temperature variations to the water-limited models in \citet{Lobo_2023}, which they attribute to less energy transport in a drier atmosphere compared to their aquaplanet cases. Our findings of more nightside cloud cover (in the less-irradiated cases; Figure \ref{fig:tsurf_maps}) also match the findings in \citet{Turbet2023}, who investigated what was required for a ``hot start'' case to condense liquid water oceans, analogous to the likely formation conditions of rocky planets. The mechanism that drives the nightside cloud is different between their study and ours. In their case, the high-altitude water vapor is an efficient absorber of incoming stellar radiation, and the resulting heating of the upper atmosphere breaks dayside convection, preventing the formation of convective clouds. Our cooler simulations have no such high-altitude water content (Figure \ref{fig:humidity_125}), and instead convection is likely inhibited by lack of available atmospheric water. 

\begin{figure}[b]
\includegraphics[width=0.45\textwidth]{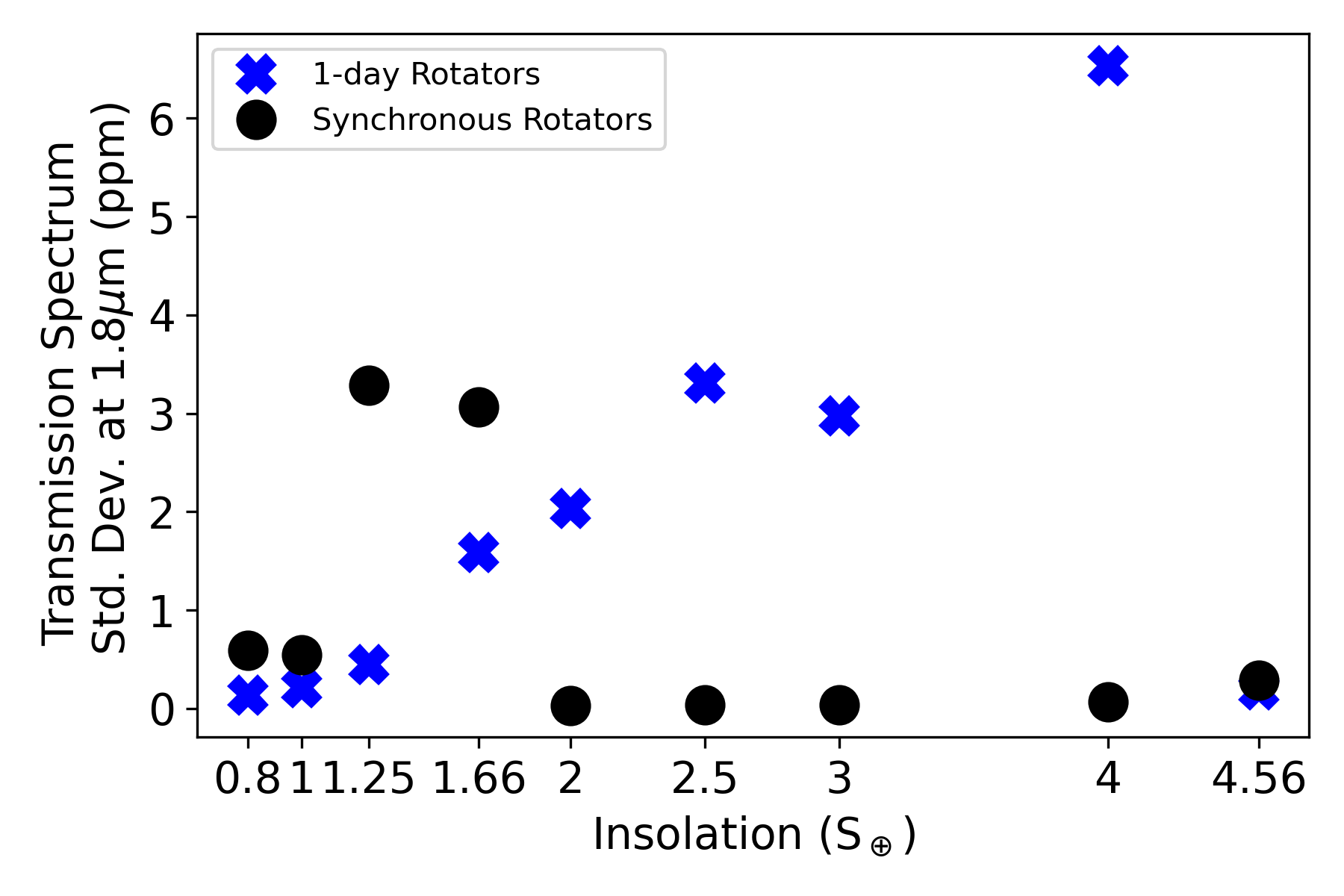}
\caption{The standard deviation of the transit depth in the 1.8~$\mu$m bandpass for monthly cadence simulations over two years. 
}
\label{fig:variability}
\end{figure}

We also investigate potential variability in our simulations, significant in light of the findings of \citet{Fauchez_2022_thai3} that inter-transit variability was larger than spectral differences between, for example, O$_2$ and H$_2$O \citep{Wunderlich2020}. Outputting the simulation instances used to generate Figures \ref{fig:spectra} at one-month intervals for two years, we measure the variation in transit depth of the 1.8$\mu$m bandpass, since it and other continuum regions are the most highly varying part of transmission spectra in the near-infrared \citep{Fauchez_2022_thai3}. We find that the standard deviation of the spectra in this region is at most around 5~ppm (Figure \ref{fig:variability}). This is roughly consistent with the amplitudes found in previous variability studies \citep[e.g.,][]{May2021, Song2021, Cohen2022}. The variability peaks for the synchronously rotating simulations in the 1.25~S$_\oplus$ and 1.66~S$_\oplus$ cases, which we observe to have tenuous, high-altitude clouds that are prone to fluctuation. Given that our simulations are fairly dry land planets without significant topography, these variable cloud effects are likely the primary driver of variability. Meanwhile, for the 1-day rotators, the variability increases with irradiation until the hottest case. This may be because the 1-day rotators maintain nonzero cloud cover until the hottest instance (Figure \ref{fig:clouds}), although we reiterate the cautions of Section \ref{subsec:limitations} around the results of these hottest instances. In any case, this peak variability amplitude of $\sim$5 and $<$1~ppm from the synchronous and 1-day rotators, respectively, in the 1.25~S$_\oplus$ case is not sufficient to completely wash out the maximum $\sim$20~ppm difference between their spectra (Figure \ref{fig:spectra}), although it could make the differences more difficult to detect. 
Overall, this work suggests that the transmission spectra of water-limited super-Earths will be fairly robust to inter-transit atmospheric variability, in line with previous work \citep[e.g.,][]{Fauchez_2022_thai3}, although the variability will make the detection of small effects like those we examine here all the more difficult. 

\subsection{The Differences We Present Are Difficult to Individually Observe}
\label{sec:observations}

We have shown above that there are visible differences in the idealized transmission spectrum of a small planet depending on whether it rotates synchronously or not. However, with real observations of an atmosphere identical to the one we model, it would be difficult or impossible to extract these differences.
Observations of small planets are already challenging. In this case, we find at best a model difference of roughly 20~ppm, a precision that could only be achieved with large numbers of JWST transits \citep[e.g.,][]{Pidhorodetska_2021}. Furthermore, the difference would require observations beyond 5~$\mu$m to break the degeneracy with a larger planetary radius (Figure \ref{fig:spectra}), but even JWST's Mid-Infrared Instrument is probably unsuitable for transmission spectroscopy \citep[though not emission, which we do not simulate here; e.g.,][]{Greene2023,Zieba2023} observations of small planets \citep{Batalha2018}. 
We do note that Figure \ref{fig:clouds} suggests that morning and evening terminators have significantly different cloud cover in the synchronous cases (especially 1.25~S$_\oplus$). In principle, this would yield differing transmission spectra for ingress and egress, but we do not calculate these explicitly given that the size of the signal would be even smaller than the already difficult-to-observe effects on the complete transmission spectrum.
In any case, the difficulty of observing these effects was to be expected given our model setup: we are considering mostly N$_2$/CO$_2$ atmospheres (i.e., high mean molecular weight). This is a fairly ``pessimistic'' (from the observational perspective) case even for a super-Earth atmosphere \citep[e.g., ][]{Pidhorodetska_2021}. 

Even in a more ``optimistic'' (i.e., lower mean molecular weight) atmosphere, this work is not sufficient to claim that planetary synchronization state can be identified from transmission spectra alone. The most significant differences that we present here are driven by clouds, but cloud formation physics is not fully understood \citep[e.g.,][]{Gao_2021_aerosols_review} and is parameterized within our model. Different and potentially more detailed parameterizations may yield different results \citep[e.g.,][]{Sergeev2020, Lefevre2022}. Thus, a ``synchronization signature'' will inevitably be degenerate with other physical processes that drive cloud formation and climate such as initial water inventory. Particularly in the case of planets with oceans included, even changing the land/ocean distribution can result in differing transmission spectra, resulting in fundamental degeneracies in planetary climate \citep{Macdonald2022, Macdonald2024}. Finally, as mentioned above, our simulations do not explore the formation of hazes, which are likely a significant opacity source in planetary atmospheres \citep[e.g.,][]{Gao_2021_aerosols_review, Cohen2024}. That said, this work provides a first pass look at how future observatories, in combination with bespoke forward models, may be able to observationally investigate planetary synchronization state. 

We must also note that ROCKE-3D is not the only GCM, and there is known to be dispersion in the results that different GCMs produce, depending on the details of their methodology. In particular, the THAI collaboration performed intercomparisons between ROCKE-3D, the Exoplanet Community Atmospheric Model (ExoCAM), The Laboratoire de Météorologie Dynamique—Generic model (LMD-G), and the Met Office Unified Model (UM). \citet{Sergeev2022_thai2} treat the moist cases among the THAI models, of which our model is most similar to their Hab 1 state. As described above, Hab 1 has significantly more water than our model (ROCKE-3D's implementation of Hab 1 features a 90~m global ocean) and less irradiated than most of the models we consider, but is otherwise similar (1 bar N$_2$ atmosphere with $\sim$400ppm CO$_2$). They find that the GCMs have broadly similar cloud behavior \citep[i.e., large cloud decks at the substellar point, following, e.g.,][]{Yang2013, Yang2014, Kopparapu2016}, but with important differences between them. ROCKE-3D produces atmospheres with the largest fractional cloud cover, while ExoCAM produces clouds with both the highest altitude and highest mixing ratio. Clouds are the main (though not only; Figure \ref{fig:spectra_cloudfree}) driver of spectral differences between our models, so a GCM like LMD-G \citep[the least cloudy of the four studied by][]{Sergeev2022_thai2} may find smaller differences between rotation states. Considering the transmission spectra resulting from the various GCMs' Hab 1 models, \citet{Fauchez_2022_thai3} find up to a 50\% spread in the number of observations required to confidently detect CO$_{2}$, which functions as a proxy for the ability to detect the presence of an atmosphere given the relative magnitude of the 4.3~$\mu$m CO$_{2}$ feature. Given the nonlinear cloud behavior we find in our simulations, the ``GCM uncertainty factor'' is relevant to our predictions as well, and similar studies with other GCMs would be needed to validate our predictions.

\subsection{Lessons for Wider Parameter Spaces}

Because of the evident non-linearity of 3D simulations (see e.g. the formation of a high-altitude cloud deck in the 1.25 S$_\oplus$ synchronous rotator case), we cannot straightforwardly ``scale up'' our simulations to make predictions for planets in other parts of parameter space, like larger sub-Neptunes, H$_2$O-rich ``water worlds,'' or hotter terrestrials that will form the majority of those observed with JWST. However, some of the physical processes we identify here are likely relevant to those planets. 

First, we describe some non-linearity in cloud formation and behavior across our simulations. Although the average percentage cloud cover follows a fairly straightforward pattern from significant clouds at the coolest temperatures to little/no cloud cover in hotter cases (Figure \ref{fig:clouds}), this does not fully capture the three-dimensional cloud behavior. The 1.25 S$_\oplus$ synchronous rotator case exemplifies this best, with high altitude clouds (Figure \ref{fig:clouds}c) that affect the behavior of the spectrum (Figure \ref{fig:spectra}, upper right panel) and climate (Figure \ref{fig:convergediags}). This result complicates ongoing efforts to understand trends in cloud/haze patterns in exoplanet atmospheres \citep[e.g.,][]{Gao_2021_aerosols_review, Yu2021_NatAs, Brande2024}, and suggests any such relationships will likely have an inherent, effectively stochastic level of scatter due to nonlinear effects like these. Despite the model limitations we describe in Section \ref{subsec:limitations}, there is no reason to think that effects of this sort are unique to our parameter space. This further suggests that 1D forward models may be insufficient for predicting an individual planet's cloud behavior. 

A related implication for our population-level understanding of exoplanet cloud behavior is what our results suggest of the behavior of condensibles other than water. 
While more work would be needed to determine how commonly the transition from condensible to too-rarefied occurs for a given species, the result that condensation produces nonlinear and inhomogeneous effects should be broadly applicable. 
Various species are thought to play the role of condensibles on hot exoplanets that they do not in our solar system, such as KCl and MgSiO$_3$, among others \citep[e.g.,][]{Gao2018,Herbort2022}. 
Although the physics and dynamics of atmospheres where these species are the dominant condensibles will be different from the simulations we present here, similar effects could play out, with limited transfer from a lower-atmosphere reservoir broadly analogous to a limited starting water inventory.
Our ultimate point is that this further motivates 3D modelling for individual planets to better understand their cloud and climate behavior. 




\subsection{Synchronization State Can Inform Sample Selection}

In addition to the implications for next-generation observations of individual small planets, where synchronization state could be a parameter inferred by spectroscopic observations, our work here has implications for current observational campaigns. In particular, efforts to infer the presence of a small planet atmosphere from JWST transmission spectroscopy requires finding signals very close to the limit of what it is possible to detect. In constructing a sample of moderately irradiated planets for observation, and subsequently attempting population-level inferences, synchronization state will play a role. Although there is no mechanism currently to determine synchronization directly, system characteristics that impact it, especially planetary multiplicity \citep{vinson_2019_spinstates, Chen2023,Shakespeare2023}, will need to be considered. 

\section{Conclusions}\label{sec:conclusions}

The key takeaways of our work are as follows:

\begin{enumerate}
    \item We created a suite of simulations of a fairly dry super-Earth exoplanet with a N$_2$/CO$_2$ atmosphere and demonstrated that, at Earth-like levels irradiation (up to $\sim$1.66~S$_\oplus$), differing planetary synchronization states produce different transmission spectra. These differences are not observed in hotter planets, although we reiterate the caveats of Section 2.2 that these simulations are in a regime of questionable validity for ROCKE-3D. 
    \item We show that the differences in transmission spectra are driven primarily by the presence of non-convective cloud decks on the terminator. The atmospheric water column profiles also drive some spectral differences at lower temperatures, before the atmosphere becomes perpetually sub-saturated. 
    \item Even though our particular simulations, which have high mean-molecular-weight nitrogen atmospheres, do not show differences at levels that could be detected with reasonable expenditures of JWST time, planets that are larger, have less dense atmospheres, or have even smaller host stars could be laboratories for this phenomenon. However, we reiterate that the differences between transmission spectra of synchronous vs. 1-day rotators should be investigated with multiple GCMs to be considered robust, given previously shown dispersion in outcomes from different models. As well, publicly available GCMs capable of handling larger, hotter planets will be important for confirming the applicability of our results outside of the specific parameter space that we explore.  
\end{enumerate}

\section{Acknowledgements}

The authors wish to thank Geronimo Villanueva and Eric Wolf for discussions and technical feedback that greatly aided the completion and quality of this manuscript. 
N.S. wishes to thank Natalie Batalha for mentorship that also greatly enhanced the quality of this manuscript. 

Support for N.S. was provided by NASA through a grant from the Space Telescope Science Institute, which is operated by the Association of Universities for Research in Astronomy, Inc., under NASA contract NAS 5-03127. 
N.S. gratefully acknowledges support from the Heising-Simons Foundation through grant 2021-3197.
C.E.H. acknowledges support through the ROCKE-3D ICAR grant.
This material is based upon work supported by NASA’S Interdisciplinary Consortia for Astrobiology Research (NNH19ZDA001N-ICAR) under award number 19-ICAR19\_2-0041. 
This work was supported by NASA's Nexus for Exoplanet System Science (NExSS) and the NASA Interdisciplinary Consortia for Astrobiology Research (ICAR). Resources supporting this work were provided by the NASA High-End Computing (HEC) Program through the NASA Center for Climate Simulation (NCCS) at Goddard Space Flight Center. 

The simulations used in this work were performed using energy-intensive supercomputing resources which have non-negligible carbon emissions and associated climate impact. In the spirit of transparency, we describe the energy usage estimates of our simulations here, and make the simulation results available publicly at doi:10.5281/zenodo.15226078. These simulations were run on the Discover cluster’s Scalable Unit 16 nodes (‘Cascade’) which have 2 24-core processors per node. Each processor has a thermal design power (TDP) of 205 W per core, meaning that one wall clock hour of operations consumes 0.41 kWh per node. Our final simulations ran for a total of 3,840 wall clock hours to reach steady state, resulting in a total power usage of 1574.4 kWh. According to the EPA’s Greenhouse Gas Equivalencies Calculator\footnote{\url{https://www.epa.gov/energy/greenhouse-gas-equivalencies-calculator}}, this translates to roughly 495 kg of CO$_{2}$ emitted.

%




\bibliography{main.bib}
\bibliographystyle{aasjournal}



\end{document}